\documentclass{svjour2}

\smartqed

\usepackage{graphicx}

\journalname{Int J Theor Phys}

\begin{document}

\title{A Dark Energy Model in Kaluza-Klein Cosmology}
\titlerunning{A Dark Energy Model ...}

\author{Utpal Mukhopadhyay \and Ipsita Chakraborty \and Saibal Ray \and A.A. Usmani}

\authorrunning{Mukhopadhyay \and Chakraborty \and Ray \and Usmani}

\institute{Utpal Mukhopadhyay \at Satyabharati Vidyapith, Nabapalli, North 24
Parganas, Kolkata 700126, West Bengal, India
\email{utpalsbv@gmail.com} \and Ipsita Chakraborty \at Department of Physics,
Adamas Institute of Technology, Barasat, North 24 Parganas, Kolkata 700126, West
Bengal, India \email{ipsita14@gmail.com} \and Saibal Ray \at Department of Physics, Government
College of Engineering and Ceramic Technology, Kolkata 700 010,
West Bengal, India \email{saibal@iucaa.ernet.in} \and A.A. Usmani \at Department of Physics, Aligarh Muslim University,
Aligarh 202002, Uttar Pradesh, India \email{anisul@iucaa.ernet.in}}

\date{Received: date / Accepted: date}

\maketitle

\begin{abstract}
We study a dynamic $\Lambda$ model with varying
gravitational constant $G$ under the Kaluza-Klein
cosmology. Physical features and the limitations of the present model
have been explored and discussed. Solutions are found mostly
in accordance with the observed features of the accelerating universe.
Interestingly, signature flipping of the deceleration parameter is noticed
and the present age of the Universe is also attainable under
certain stringent conditions. We find that the time variation of
gravitational constant is not permitted without vintage $\Lambda$.

\end{abstract}

\section{Introduction}

The cosmological picture of an accelerating Universe which came
into limelight through seminal papers of Perlmutter et al.
\cite{Perlmutter1998} and Riess et al. \cite{Riess1998} is, at
present, a well established scientific truth. The accelerating
agent is termed as dark energy and the vintage cosmological term
$\Lambda$ is a favourite choice of the researchers working with
 Lambda-dark energy models as the dark energy representative.
Although Einstein \cite{Einstein1917} introduced $\Lambda$ as a
constant in his field equations, due to Cosmological Problem and
Coincidence Problem, now-a-days in most of the cases it is
considered as a dynamical quantity \cite{Overduin1998}. On the
other hand, the gravitational constant $G$ is, in general, taken
as a constant. However, sufficient amount of works related to
variability of $G$, both at theoretical and experimental front,
are also available in literature \cite{Ray2007a,Mukhopadhyay2010}.

In the third decade of the previous century, Kaluza
\cite{Kaluza1921} and Klein \cite{Klein1925} attempted to unify
electro-magnetic force with gravitational force which resulted in
the development of Kaluza-Klein (KK) theory. In KK approach, an
extra dimension, viz. fifth dimension was introduced for coupling
the two forces mentioned earlier. Chodos and Detweiler
\cite{Chodos1980} have shown in their five dimensional model that
the extra dimension contracts due to cosmic evolution. According
to Guth \cite{Guth1981} and Alvarez and Gavela \cite{Alvarez1983},
production of huge entropy due to the presence of an extra
dimension can solve the flatness and horizon problems without
invoking the idea of inflation. So, five dimensional model in the
framework of KK theory has been successful in addressing some of
the problematic issues of Big Bang cosmology and other realm of physics
(for a critical review on the KK theory one can look at the new-born article
by Wesson \cite{Wesson2015}).

A number of works are available in the literature where one or
both of the cosmological constant $\Lambda$ and the gravitational
constant $G$ are assumed to be variable. Pradhan et al.
\cite{Pradhan2008} and Ozel et al. \cite{Ozel2010} worked in the
framework of KK cosmology with variable $\Lambda$ and constant $G$
while Mukhopadhyay et al. \cite{Mukhopadhyay2010} performed an
$n$-dimensional investigation with both $\Lambda$ and $G$ as
variables. Sharif and Khanum \cite{Sharif2011} and more recently
Oli \cite{Oli2014} have worked with KK cosmological models by
considering both $\Lambda$ and $G$ as variables. A special note
to the work of Sharif and Khanum \cite{Sharif2011} is
that they have investigated the effect of $\Lambda =
\epsilon H^2$ model (where $\epsilon$ is a parameter) in KK
cosmology with variable $\Lambda$ and $G$. Ray et al.
\cite{Ray2007a} showed the equivalence of $\Lambda \sim H^2$,
$\Lambda \sim \ddot a/a$ and $\Lambda \sim \rho$ models in the
four dimensional FRW spacetime. So, the major motivations of the
present work is to explore those features which
were not touched upon in the work of Sharif and Khanum~\cite{Sharif2011}.
Although generalized energy conservation
law for variable $\Lambda$ and $G$ models have been derived by
Shapiro et al. \cite{Shapiro2005} and Vereschagin et al.
\cite{Vereschagin2006} in two different ways, Shapiro et al.
\cite{Shapiro2005} have mentioned in their work that without any
loss of generality usual energy conservation law can be retained
for variable $\Lambda$ and variable $G$ models. Beesham
\cite{Beesham1993} has supported this idea of retention of usual
energy conservation law by stating that the usual energy
conservation law can be used within a simple framework of variable
$\Lambda$ and $G$
\cite{Lau1985,Abdel-Rahman1990,Sistero1991,Berman1991,Kalligas1992}.

The paper is organized as follows. Section 2 deals with field
equations and their solutions; physical features explored in the
work are presented in Section 3 while some discussions are done in
Section 4.

\section{Field equations and their solutions}

The metric of KK cosmology is given by
\begin{eqnarray}
ds^2 = dt^2 - a^2(t) \left[ \frac{dr^2}{1 - kr^2} + r^2 d\Omega^2 + (1 - kr^2)d\psi^2\right]
\end{eqnarray}
where $d\Omega^2 = d\theta ^2 + sin^2 \theta d\phi^2$.

Perfect fluid energy-momentum tensor takes the usual form as
\begin{eqnarray}
T_{ij} = (p + \rho) u_{\mu}u_{\nu} - pg_{\mu\nu},
\end{eqnarray}
where $\mu$, $\nu$ = $0$, $1$, $2$, $3$, $4$ and $u_{\mu}$ are five-
velocity satisfying $u_{\mu}u_{\nu}$ = $1$, $\rho$ is the energy
density and $p$ is the pressure of the cosmic fluid.

The Einstein field equations are given by
\begin{eqnarray}
R_{ij}- \frac{1}{2}Rg_{ij} + \Lambda g_{ij} = 8\pi GT_{ij},
\end{eqnarray}
where the terms used have their usual meanings.

Using (1) - (3) we have the following two equations
\begin{eqnarray}
8\pi G\rho + \Lambda = 6\left(\frac{\dot a^2}{a^2} + \frac{k}{a^2}\right),
\end{eqnarray}

\begin{eqnarray}
8\pi G p - \Lambda = -3\left(\frac{\ddot a}{a} + \frac{\dot a^2}{a^2}
 + \frac{k}{a^2}\right).
\end{eqnarray}

For a flat Universe, $k=0$ and hence Eqs. (4) and (5) reduce to
\begin{eqnarray}
8\pi G\rho + \Lambda = 6\frac{\dot a^2}{a^2} = 6H^2,
\end{eqnarray}

\begin{eqnarray}
8\pi G p - \Lambda = - 3\dot H - 6H^2,
\end{eqnarray}
where $H = \dot a/a$ is the Hubble parameter.

The continuity equation yields
\begin{eqnarray}
\dot \rho + 4H(p+\rho) = 0,
\end{eqnarray}
while the equation of state is
\begin{eqnarray}
p = \omega \rho,
\end{eqnarray}
where $\omega$ is the barotropic index.

Let us use the ansatz $\Lambda = \beta (\ddot a/a) = \beta (\dot
H+ H^2)$, where $\beta$ is a parameter. This type of model was
dealt by  Overduin and Cooperstock \cite{Overduin1998} and
Arbab \cite{Arbab2003a,Arbab2003b,Arbab2004} to account for the
accelerating phase of the present Universe.

Then equation (6) becomes
\begin{eqnarray}
\beta \dot H + (\beta - 6)H^2 = -8\pi G\rho.
\end{eqnarray}

Also equation (7) reduces to
\begin{eqnarray}
(\beta - 3)\dot H + (\beta - 6)H^2 = 8\pi G p.
\end{eqnarray}

Dividing (11) by (10) and then simplifying we get
\begin{eqnarray}
\frac{\dot H}{H^2} =
-\frac{(1+\omega)(\beta-6)}{(\beta+\beta\omega-3)},
\end{eqnarray}
which on integration yields the solution set as
\begin{eqnarray}
a(t) = Ct^{\frac{\beta+\beta\omega-3}{(1+\omega)(\beta-6)}},
\end{eqnarray}

\begin{eqnarray}
H(t) = \frac{\beta+\beta\omega-3}{(1+\omega)(\beta-6)t},
\end{eqnarray}

\begin{figure}
\includegraphics[scale=.4]{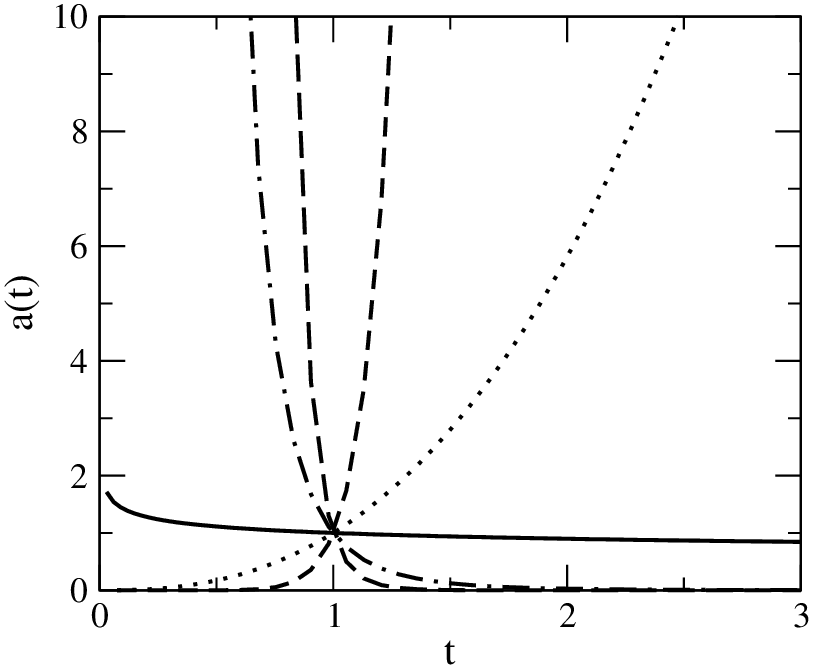}
\caption{Plot for variation of scale factor, $a$, with respect to
age of the Universe, $t$, as given by equation (13). The solid,
dotted, dashed, long-dashed and dot-dashed curves represent
$\omega =0, -0.7, -0.9, -1.1$ and $-1.3$, respectively.}
\end{figure}

\begin{eqnarray}
\Lambda (t) =
\frac{3\beta(1+2\omega)(\beta+\beta\omega-3)}{(1+\omega)^2(\beta-6)^2t^2},
\end{eqnarray}

\begin{eqnarray}
\rho(t) = C_1t^{\frac{-4(\beta+\beta\omega-3)}{(\beta-6)}},
\end{eqnarray}

\begin{eqnarray}
G(t) =\frac{3}{8\pi C_1}
\frac{\beta+\beta\omega-3}{(1+\omega)^2(\beta-6)}
t^\frac{2\beta(1+2\omega)}{\beta-6},
\end{eqnarray}

\begin{eqnarray}
q =  - \frac{a\ddot a}{\dot a^2}
= -{\frac{3 + 6\omega}{\beta +\beta\omega -3}}.
\end{eqnarray}

To obtain the present-day feature of the dust filled Universe one
should put the constraint $\omega=0$ in the general form of Eqs.
(13)-(18).

\section{Physical Features}

The physical features of the present model in the framework of
Kaluza-Klein theory are buried in its solution set through
Eqs.~(13) - (17). We choose here the numerical value of $\beta$
from the range $3.417 \leq \beta \leq 4.674$ in its lower limit
i.e. $\beta=3.4$~\cite{Arbab2004,Ray2007b}, which represents the
observed accelerated expansion, and then plot these solutions for
different values of $\omega=0.0, -0.7, -0.9, -1.1$ and $-1.3$ as
represented by solid, dotted, dashed, long-dashed and dot-dashed
curves in Figs.~1-5. We take here, for simplicity, $C=C_1=1$ in
Eqs. (14), (16) and (13), where they appear as a multiplier or
divider.

The obvious physical features are discussed below:

(i) We find the condition, $\beta \neq 6$, for any physical validity of the
solutions.

(ii) For the values, $\beta=0$ or $\omega=-0.5$ or
$\beta=3/(1+\omega)$, $\Lambda$ becomes zero. Here, the last two conditions,
$\omega=-0.5$ and $\beta=3/(1+\omega)$, yield the unphysical situation,
$\beta=-6$.

(iii) Also, along with $\Lambda$ the Hubble parameter reduces to zero
with $\beta=3/(1+\omega)$ while both the scale factor and the
matter-energy density become constant. None of these features
support the present cosmological scenario.

(iv) With $\beta=0$, $G$ becomes constant.
Thus the present model does not permit time variation of
$G$ in the absence of the cosmological parameter $\Lambda$.

\begin{figure}
\includegraphics[scale=.4]{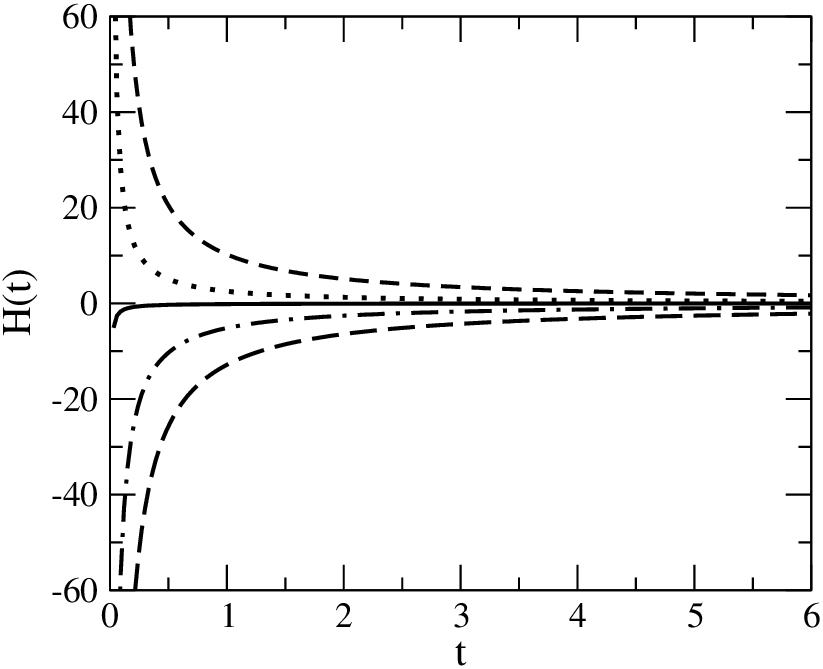}
\caption{The description of various curves for Hubble parameter is
the same as in Fig. 1.}
\end{figure}

(v) We also find $t=1/2H$ from Eq.~(14) with $\beta=\omega=0$. With
the present value of the Hubble parameter as
$72~kms^{-1}Mpc^{-1}$, the age of the Universe comes out to be $7$
Gyr which is much below the accepted range of $13.2$
to $14$ Gyr (see Table 2 of the Ref. \cite{Ray2007b}).
If we want to retain the non-zero values of $\beta$ and $\omega$
such that $3.4$ and $-0.36$ respectively then the above value of the age,
viz. $7$ Gyr can be obtained. However, to attain the modern-day accepted
value $\sim 14$ Gyr one has to put $\omega = -0.5$. Thus it is observed
that with proper tuning of $\beta$ one may recover the age of the
Universe for any value of $\omega$. In this regard we would also like to
make a general comment that all the above discussion justify the necessity
of the inclusion of $\Lambda$ in the field equations.

(vi) We also find inverse square law between time and density,
$\rho\sim t^{-2}$, for $\omega=0$. This
relationship of $\rho$ with cosmic time was obtained earlier by
Ray et al. \cite{Ray2007b,Ray2007c} in four dimensional model with
$\Lambda$.

(vii) A negative value of the deceleration parameter, $q$,
signifies the present accelerating phase. For the given value
$\beta=3.4$, the deceleration parameter $q$ (Eq.~18) flips the
sign and becomes positive if $-0.118>\omega> -0.5$. For the dust
filled case ($\omega=0$), $q$ flips the sign if $\beta <3$.  A
significant aspect of the evolution of the Universe is the flip
over from a previously decelerating phase to the present
accelerating one \cite{Riess2001,Padmanabhan2003,Amendola2003}.
So, the deceleration parameter $q$ must show a signature flipping
to indicate that turnover and hence evolution of the Universe.

\begin{figure}
\includegraphics[scale=.4]{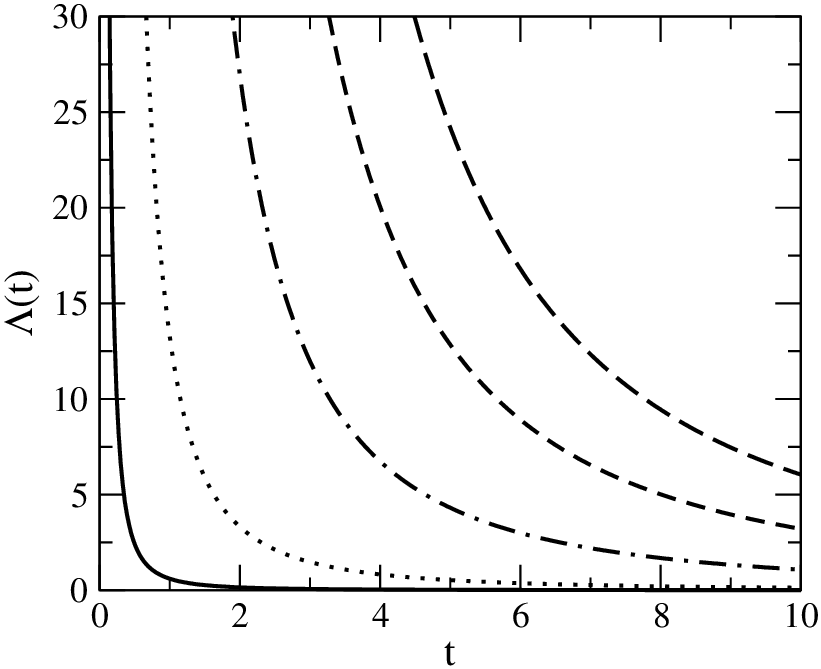}
\caption{The description of various curves for Cosmological
parameter is the same as in Fig. 1.}
\end{figure}

(viii) All the plots for $a(t)$, $H(t)$, $\Lambda (t)$, $\rho (t)$ and $G(t)$ are
found sensitive with the variation of $\omega$ as it stands in the
exponent of time and also due to the $1+\omega$ factor in the denominator.

There is a little variation for the scale factor, $a(t)$, at $t=1$
irrespective of the value of $\omega$. However, there is a large variation in
the values of $a(t)$ with respect to time. For the Hubble parameter, we find
large or small value depending on $\omega$, which tend to zero with increasing
time.

The $\Lambda$ parameter is found to be always positive for the given
$\beta$ and $\omega$ and thus acts as an dark energy agent of the
accelerating scenario of the Universe.

Like $a(t)$, there is not much variation in the density,
$\rho$, at $t=1$. For $\omega$=0, density is found increasing with time.
However, it shoots up at smaller time for non-zero values of $\omega$.
This type of feature has been observed by Ray et al. \cite{Ray2007b,Ray2009}
for which they made a comment that idea of inflation is inherent in the
phenomenological model  $\Lambda = \beta (\ddot a/a) = \beta (\dot
H+ H^2)$.

In the present investigation the gravitational parameter $G$ varies with time
and is also found sensitive with $\omega$.

\begin{figure}
\includegraphics[scale=.4]{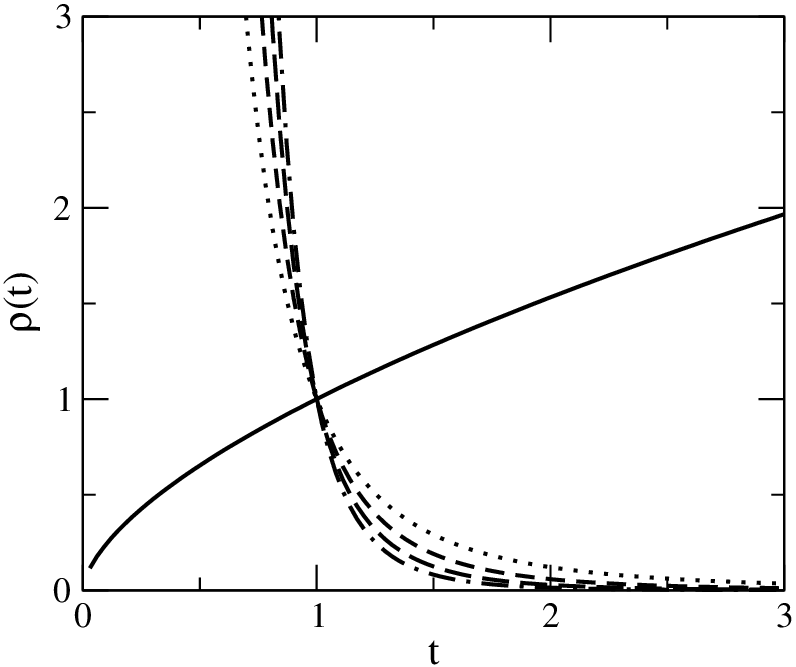}
\caption{The description of various curves for the cosmic
matter-energy density parameter is the same as in Fig. 1.}
\end{figure}

\section{Conclusions}

In the present work we study cosmic evolution  under the framework of Kluza-Klein cosmology.
The emphasis has been on dynamic $\Lambda$ model with varying gravitational
constant. We present our solutions for the scale factor, Hubble parameter,
gravitational constant and density with respect to time $t$. These are also
plotted for various values of equation of state parameter,
$\omega$. Interesting physical features of these solutions have been
summarized and discussed. Solutions are mostly found in accordance of
the observed features of the Universe. We find that the time variation
of the gravitational constant is not permitted without inclusion of $\Lambda$.
Signature flipping of the deceleration parameter is attainable. Besides,
some previous results are also recovered in the present work. For instance,
relationship of both the matter-energy density \cite{Ray2007b,Ray2007c} and gravitational
constant \cite{Ray2007a} can be recovered from Eqs. (16) and
(17) respectively. The present work therefore can be regarded as a continuation of two previous
works \cite{Ray2007a,Mukhopadhyay2010} related to variation of the
gravitational constant $G$ and also confirmation of some of the earlier results \cite{Ray2007b,Ray2007c}.

However, in the present phenomenological investigation of simple
and elegant type some of the crucial issues have not been
addressed, such as: {\it Firstly}, calculation of the evolution of
linear perturbations in this model. This point is particularly
important, as the expansion laws are easy to match with
observations, and distinguishing between models requires
information on the perturbations. {\it Secondly}, one can raise a
simple question such that where is the fifth dimension now?
Actually Eq. (1) suggests that the fifth dimension is large and is
around us. Therefore, there must be a mechanism to make it small
as modern physicists worry a lot about how to make the
extra-dimensions compactified and one can again look at the review
on KK theory by Wesson \cite{Wesson2015} to get a possible way out
from the problem. All these aspects can, for the time being, be
left as future efforts.

\begin{figure}
\includegraphics[scale=.4]{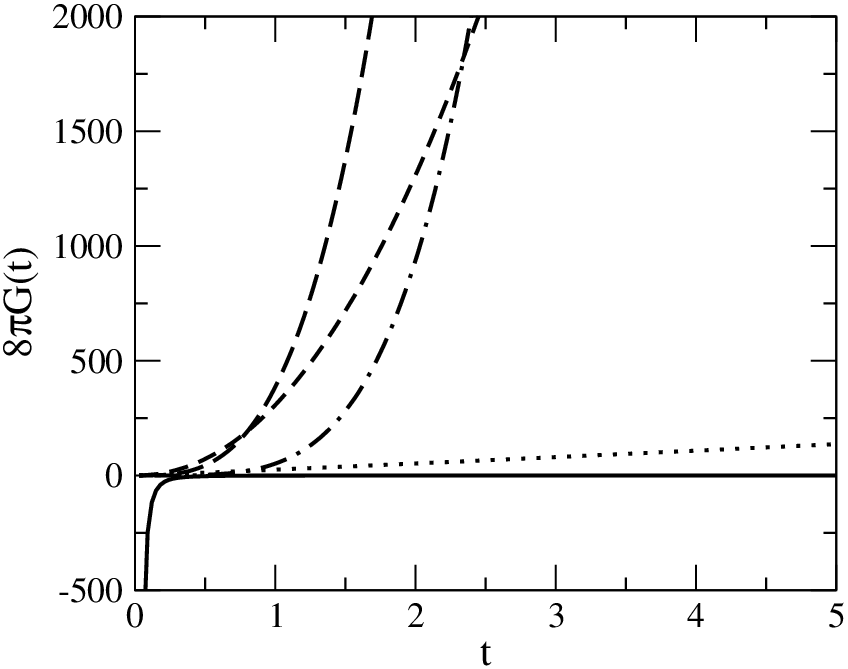}
\caption{The description of various curves for gravitational
parameter is the same as in Fig. 1.}
\end{figure}

\section*{Acknowledgments}
SR and AAU are thankful to the authority of Inter-University
Centre for Astronomy and Astrophysics, Pune, India for providing
Visiting Associateship under which a part of this work was carried
out.

\end{document}